\documentclass{article}

\usepackage{arxiv}

\usepackage[utf8]{inputenc} 
\usepackage[T1]{fontenc}    
\usepackage{hyperref}       
\usepackage{url}            
\usepackage{booktabs}       
\usepackage{amsfonts}       
\usepackage{nicefrac}       
\usepackage{microtype}      
\usepackage{lipsum}		
\usepackage{graphicx}
\usepackage{doi}

\usepackage[authoryear,round]{natbib}

\usepackage{tabularx}
\usepackage{booktabs}
\usepackage{array}
\usepackage{tikz}
\usetikzlibrary{arrows.meta, positioning}
\usepackage{amsmath}

\hypersetup{
    colorlinks=true,
    linkcolor=black,
    citecolor=black,
    urlcolor=black,
    filecolor=black
}

\title{Technohybridism: Recursive Semantic Normalization in AI‑Mediated Societies}

\author{
    \href{https://orcid.org/0000-0002-4717-031X}{\includegraphics[scale=0.06]{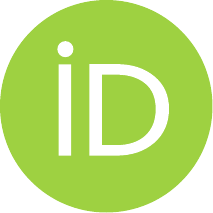}\hspace{1mm}Charalampos M. Liapis}
    \\
    Department of Mathematics, University of Patras, Patras 26504, Greece\\
    Department of Philosophy, University of Ioannina, Ioannina 45110, Greece\\
    Computer Technology Institute and Press "Diophantus", Patras 26504, Greece\\
    \texttt{c.liapis@ac.upatras.gr}
    \And
    \href{https://orcid.org/0000-0002-1576-5300}{\includegraphics[scale=0.06]{orcid.pdf}\hspace{1mm}Epameinondas Panagopoulos}
    \\
    Department of Educational Sciences and Social Work, University of Patras, Patras 26504, Greece\\
    Department of Primary Education, University of Ioannina, Ioannina 45110, Greece\\
    Computer Technology Institute and Press "Diophantus", Patras 26504, Greece\\
    \texttt{epanagopoulos@upatras.gr}
}

\date{}

\renewcommand{\shorttitle}{Technohybridism: Recursive Semantic Normalization in AI‑Mediated Societies}

\renewcommand{\headeright}{Preprint}
\renewcommand{\undertitle}{}

\newcommand{\authors}{Liapis, C.M.; Panagopoulos, E.}

\hypersetup{
pdftitle={Technohybridism: Recursive Semantic Normalization in AI‑Mediated Societies},
pdfsubject={cs.CY, cs.AI, cs.HC},
pdfauthor={Charalampos M. Liapis, Epameinondas Panagopoulos},
pdfkeywords={Generative AI, Large language models, Recursive semantic normalization, Technohybridism, Social epistemology, Semantic norms},
}

\begin{document}
\maketitle

\begin{abstract}
	Language-generating AI systems are no longer only objects of linguistic interaction: their outputs increasingly re-enter the communicative and institutional practices from which their training material was drawn. Drawing on social theory, social epistemology, and the philosophy of language, this article develops a sociotechnical account of that re-entry. We introduce \emph{recursive semantic normalization}: the process through which machine-generated formulations, once recurrently taken up, evaluated, reused, and institutionally retained, participate in the reproduction and transformation of semantic regularities. Under the heading of \emph{technohybridism}, we argue that the relevant unit of analysis is the \emph{hybrid semantic system}, a recursively coupled configuration of human practices, computational generation, and institutional uptake. First, we distinguish the concept from neighbouring accounts, including model collapse, looping effects, machine habitus, machine culture, and the epistemology of AI testimony. Second, we develop an analytical model that separates generation, selection, uptake, and persistence and distinguishes machine-attributable change from change that would have occurred anyway. Third, we analyse uptake as routinized, institutionally diffused, and socially stratified practice and state conditions under which machine-derived regularities become norms. Fourth, we identify social-epistemic consequences: the laundering of provenance that weakens epistemic vigilance, the erosion of the cognitive diversity on which collective inquiry depends, and the narrowing of shared hermeneutical resources. An evidential ladder for empirical research follows. The account attributes no normative authority to machine outputs; what changes is the field within which human and institutional uptake operates.
\end{abstract}


\section{Introduction}

In AI-mediated societies, linguistic interaction is increasingly routed through systems that do not merely transmit language but generate it. Large language models (LLMs) are trained on large corpora of human-produced text and generate context-sensitive continuations by exploiting statistical regularities in language \citep{mikolov2013distributed, vaswani2023attentionneed,brown2020languagemodelsfewshotlearners}. Their outputs are read, edited, accepted, rejected, cited, and reused in scientific writing, education, journalism, administration, software development, and everyday exchange. This circulation is no longer a matter of conjecture. Population-level estimates indicate that at least 13.5\% of biomedical abstracts published in 2024 were processed with LLMs \citep{doi:10.1126/sciadv.adt3813}, that between 6.5\% and 16.9\% of the text of peer reviews submitted to several major AI conferences in 2023-2024 may have been substantially modified by LLMs \citep{pmlr-v235-liang24b}, and that by late 2024 roughly 14\% of the text of United Nations press releases and up to 24\% of that of corporate press releases was LLM-assisted \citep{LIANG2025101366}. A study of spoken language, so far available as a preprint, reports that words characteristic of LLM output have become more frequent in unscripted human speech since the release of ChatGPT and that brief exposure to a chatbot's word choices induces their adoption \citep{yakura2026empiricalevidencelargelanguage}.

These findings concern who, or what, takes part in producing text, not yet what words mean. They nonetheless make pressing a question that is usually overshadowed by debates about whether LLMs understand, refer, or merely imitate \citep{bender-koller-2020-climbing,10.1093/oso/9780192894724.001.0001,10.1162/coli_a_00522}. Whatever the answer to those debates, the linguistic regularities learned from historically situated human discourse can be computationally regenerated and reintroduced into the practices from which they were drawn. The question is then not only what a model's outputs mean, but what happens to the social organization of meaning when such outputs become recurrent inputs into it. For the social sciences, the question is pressing because language is not only a means of communication but the medium through which social reality is typified, institutions are legitimated, and knowledge is shared \citep{berger1966social,habermas1985theory,bourdieu1991language}.

We call this process \emph{recursive semantic normalization} (RSN): the process through which machine-generated linguistic material, derived from historically structured human language, is reintroduced into social practices and, through recurrent uptake, selection, correction, reuse, or institutionalization, participates in the stabilization, reproduction, and transformation of semantic regularities. The claim is not that machine outputs possess normative authority in their own right, nor that statistical regularity amounts to normativity. It is that machine-generated formulations can become part of the human practices through which expressions acquire salience, repeatability, intelligibility, and, under some conditions, normative significance. We develop this claim under the heading of \emph{technohybridism}. Under conditions of sustained computational mediation, we argue, the processes through which meanings are socially stabilized should be analysed neither solely at the level of individual cognition nor exclusively within human interaction, but at the level of \emph{hybrid semantic systems}: recurrent configurations in which human practices, computational generation, and institutional uptake jointly shape which linguistic forms become available and which become stable.

The recursion at issue is primarily social. It differs from the technical recursion studied under the name of model collapse, in which generative models are retrained on their own outputs \citep{shumailov2024aimodels}. The social recursion operates whether or not any model is retrained, and it is upstream of the technical one, because it determines what future corpora will contain. The relevant sequence is therefore not simply human language in, machine language out, but a relation in which machine-generated language becomes part of the linguistic environment subsequently encountered by human users, institutions, and, potentially, future computational systems.

The central research question is accordingly: \emph{under what conditions can machine-generated linguistic material become part of the social processes through which semantic regularities are stabilized, reproduced, or transformed, and what follows when it does?} Several literatures address parts of this question. Philosophy of language asks whether model outputs are meaningful; social epistemology asks whether reliance on them is justified; machine learning studies recursive training; research on AI-mediated communication documents how assisted writing alters messages and attitudes; the sociology of algorithms theorizes feedback between machine outputs and social dispositions; and cultural evolution examines how conventions spread and change. None of them, we argue, specifies the mechanism by which machine-generated formulations pass from transient output to socially consequential semantic regularity, the social conditions under which this passage occurs, or what it implies for the shared resources with which communities form and criticize beliefs.

The article makes five contributions. The first is conceptual: it defines RSN with explicit threshold and boundary conditions and distinguishes it systematically from seven neighbouring accounts (Sect. 2, Table 1). The second is analytical: it specifies the mechanism as a normative-statistical recursion and develops it into an analytical model that identifies what empirical work must establish and distinguishes machine-attributable linguistic change from change that would have occurred anyway (Sect. 3). The third is sociological: drawing on accounts of social norms, practice theory, institutional theory, and Bourdieu's analysis of linguistic capital, it treats uptake as routinized, institutionally diffused, and socially stratified practice, and states conditions under which machine-derived regularities become semantic norms (Sects. 4-5). The fourth is social-epistemic: it analyses how RSN bears on epistemic vigilance, cognitive diversity, hermeneutical resources, and the form of trust involved in reliance on machine-generated language (Sect. 6). The fifth is methodological: it derives an evidential ladder, from lexical to conceptual levels, with indicators, research designs, and conditions under which the account would be disconfirmed (Sect. 7). Section 6 also traces semantic and historical consequences; Section 8 answers objections; Section 9 draws out implications and limitations; Section 10 concludes.

The paper is conceptual. It proceeds by conceptual analysis and theoretical synthesis, using empirical studies of LLM-assisted writing and AI-mediated communication as constraints on, and illustrations of, the framework rather than as evidence for its full scope. Its concern is sociotechnical rather than semantic in the narrow sense: not what meanings are as abstract entities, but how language-generating systems participate in the social organization through which public meanings are produced, circulated, stabilized, and revised, and how this organization conditions what communities can know and contest.

\section{Positioning: what neighbouring accounts leave open}

\subsection{Meaning, language models, and the social location of semantic norms}

Most philosophical work on LLMs and language asks whether model outputs are meaningful, whether their words refer, and whether models understand. \cite{bender-koller-2020-climbing} argue that a system trained only on linguistic form cannot learn meaning. \cite{10.1162/coli_a_00522}, drawing on externalist metasemantics, argue that the words of language models may nonetheless refer by virtue of the models' place in the linguistic community from which their training data derive. \cite{10.1093/oso/9780192894724.001.0001} and \cite{Grindrod2024} defend related accounts on which outputs can have content without the system having mental states, while \cite{Floridi2023} characterizes generative models as agency without intelligence and \cite{10.1145/3624724} cautions against anthropomorphic descriptions of their capacities. Work in the distributed-language tradition shifts attention from the system to the interaction. \cite{10.1007/s00146-025-02541-1} argues that LLM outputs persuade through familiarity and that human languaging, not the model, adds meaning to them; \cite{GahrnAndersen2026} proposes the notion of enlanguaged cognition to capture how dialogical and practically open LLM-based systems shape cognition and coordination beyond single situations; and \cite{Arif2026} describes the model's internal world as a "simulated social" constructed from statistical patterns in language data. Studies of public discourse, finally, examine how LLMs are themselves framed, for instance as mind-like agents in news media \citep{NguyenWischnewski2026}, framings that plausibly affect the authority users attribute to model outputs and hence how readily they take them up.

The externalist line of argument carries an implication that this debate rarely draws. If, as \cite{Putnam1975} and \cite{Burge1979} argued, the content of an expression depends partly on a community's practices, on a division of linguistic labour and on deference to communal usage, then many meaning-relevant facts are social facts, and a technology that participates in those practices can alter them. The debate's focus on whether models \emph{acquire} meaning from the community leaves open the converse question: whether, and how, the community's semantic practices are modified once model outputs re-enter them. RSN addresses that converse question. It is neutral on whether model outputs have content of their own; it requires only that they are taken up as meaningful by human interpreters.

\subsection{Recursion in training: model collapse}

\cite{shumailov2024aimodels} show that training generative models indiscriminately on model-generated data causes the tails of the original distribution to disappear. \cite{alemohammad2023selfconsuminggenerativemodelsmad} describe "self-consuming" loops in which, without sufficient fresh real data in each generation, quality or diversity progressively declines, and \cite{gerstgrasser2024modelcollapseinevitablebreaking} report that collapse can be avoided when synthetic data accumulate alongside, rather than replace, real data. This literature concerns a closed technical loop; its object is a model's output distribution, and its evaluative vocabulary, of degradation and collapse, is that of model quality. RSN concerns an open social loop whose object is the linguistic environment of a community, and in which every machine-derived contribution passes through human or institutional selection before it persists. The two loops are connected, since what socially persists becomes future training data, but the social loop can operate, and can have consequences, without any retraining.

\subsection{AI-mediated communication and homogenization}

Research on AI-mediated communication studies interpersonal communication in which an intelligent agent modifies, augments, or generates messages on a communicator's behalf \citep{10.1093/jcmc/zmz022}. Several findings bear directly on RSN. Predictive text leads writers to produce shorter and more predictable image captions \citep{10.1145/3377325.3377523}. Co-writing with a language model configured to favour one side of a debate shifts both the arguments people write and the attitudes they subsequently report \citep{10.1145/3544548.3581196}. AI-generated reply suggestions shift the emotional tone of conversations and affect how partners perceive one another \citep{Hohenstein2023}. Writing with an instruction-tuned model reduces the diversity of content across writers \citep{PadmakumarHe2024}; access to generative AI ideas raises the rated creativity of individual stories while making the stories more similar to one another \citep{doi:10.1126/sciadv.adn5290}; ChatGPT-assisted ideation yields less semantically distinct ideas across users \citep{10.1145/3635636.3656204}; and AI writing suggestions move Indian participants' writing toward Western styles \citep{10.1145/3706598.3713564}. Syntheses now argue that LLMs homogenize human expression and thought \citep{SOURATI2026805}.

These studies document effects of exposure and assistance, usually within single sessions or bounded samples. What they do not provide, and mostly do not aim to provide, is an account of when such effects aggregate into the semantic regularities of a community, and when those regularities acquire normative standing. RSN is intended to supply this link between session-level effects and community-level semantic structure.

\subsection{Recursivity in the sociology of algorithms}

Sociology has long theorized loops between classifications and what they classify. Hacking's "looping effects of human kinds" describe how classifying people changes the people classified, who in turn change the classification \citep{10.1093/acprof:oso/9780198524021.003.0012}. \cite{10.7551/mitpress/9780262134606.001.0001} shows that financial models can be performative, shaping the markets they describe. \cite{FourcadeJohns2020} analyse the recursivity of social and machine learning, in which human and machine dispositions feed back into one another; \cite{Airoldi2022} theorizes a "machine habitus" through which algorithms trained on socially structured data reproduce social structure and, through their outputs, shape the dispositions of users; \cite{10.7551/mitpress/14189.001.0001}, from a Luhmannian perspective, treats algorithms as participants in "artificial communication"; and \cite{CouldryHepp2017} describe the deep mediatization of social reality.

RSN belongs to this family but differs in three respects. Its object is neither persons nor markets nor rankings but semantic regularities, the patterns of expression, distinction, and inference through which contents become publicly articulable. Its \emph{channel} is not the classification, prediction, or sorting of pre-existing items but the generation of new candidate formulations that did not exist before they were proposed. And its \emph{normative locus} is specific: machine outputs enter practices already governed by norms of interpretation and correction, so that the question of how a regularity becomes a norm, which these accounts do not develop for semantic regularities, becomes central.

\subsection{Language change, cultural evolution, and machine culture}

Language change has been analysed as an invisible-hand process, the unintended aggregate outcome of intentional communicative acts \citep{Keller1994}, and as an evolutionary process of variation and selection among linguistic replicators \citep{Croft2000}. Computational studies measure semantic drift from diachronic corpora \citep{hamilton-etal-2016-cultural}, and psycholinguistic work shows that diachronic change is constrained by how people use and learn language \citep{LiSiew2022}. Research on cultural evolution examines how norms and conventions emerge, spread, and change through interaction, transmission, and selection \citep{AmatoEtAl2018,https://doi.org/10.1111/cogs.13113}, and how variants converge on "attractors" through biased transformation \citep{Sperber1996}. \cite{BrinkmannEtAl2023} extend this programme to "machine culture", in which machines generate, transmit, and select cultural variants, and \cite{doi:10.1126/science.adt9819} argue that large models are best understood as cultural and social technologies, comparable to writing, print, markets, and bureaucracies.

RSN does not replace these accounts. Its narrower claim is that, under generative AI, machine-generated linguistic material becomes one of the inputs into these processes. In Keller's terms, RSN describes an invisible-hand process with a new kind of contributor; in the terms of machine culture, it concerns machine-generated variation subject to human and institutional selection in the specific domain of semantic regularity. What it adds is a specification of conditions, a way of separating machine-attributable change from ordinary change, and an analysis of when selected variants become norms. Table 1 summarizes these contrasts, together with those drawn in Sect. 2.6.

\subsection{Social epistemology: from testimony to shared epistemic resources}

Social epistemology studies how knowledge depends on others: on testimony \citep{10.1093/0198235518.001.0001}, on epistemic dependence on experts whose reasons the recipient cannot check \citep{Hardwig1985}, and on the social practices and institutions through which beliefs are formed and evaluated \citep{Goldman1999}. Two of its results matter here. First, the epistemic performance of communities depends on their diversity. A division of cognitive labour that sustains different approaches \cite{Kitcher1990}, the presence of perspectives able to criticize one another \citep{Longino1990}, and a transient diversity of opinion that prevents premature convergence \citep{Zollman2010} all improve collective inquiry, and groups of diverse problem-solvers can outperform groups of individually more able but similar ones \citep{HongPage2004}. Second, the reliability of testimony depends on mechanisms of epistemic vigilance through which hearers calibrate their trust to the competence and benevolence of identifiable sources \citep{SperberEtAl2010}.

Social-epistemological work on AI has mostly asked a question modelled on testimony: can AI outputs be trusted, or only relied upon, and when is reliance on them justified? AI has been analysed as an epistemic technology, designed and deployed for epistemic tasks \citep{Alvarado2023}, and it has been argued that AI systems can be relied on but not trusted in the full interpersonal sense \citep{Ryan2020}. RSN shifts the object of analysis. Its concern is not whether individuals form justified beliefs from particular outputs, but how machine-generated material alters the shared resources, the formulations, distinctions, and framings, with which a community forms, expresses, and criticizes beliefs at all. \cite{Levy2022} has argued that many bad beliefs are rational responses to degraded epistemic environments, shifting attention from individual virtue to the environment in which belief formation takes place. RSN specifies one mechanism through which the epistemic environment of a community is altered, and Section 6.2 shows that it bears directly on vigilance and on diversity.

\begin{table}[htbp]
    \caption{Recursive semantic normalization (RSN) and neighbouring concepts. The comparison concerns the analytical focus of each concept, not the full scope of the literatures cited.}
    \label{tab:rsn}
    \centering
    \small
    \setlength{\tabcolsep}{3pt}
    \renewcommand{\arraystretch}{1.2}
    \begin{tabularx}{\textwidth}{
        >{\raggedright\arraybackslash}p{0.20\textwidth}
        >{\raggedright\arraybackslash}p{0.15\textwidth}
        >{\raggedright\arraybackslash}p{0.17\textwidth}
        >{\raggedright\arraybackslash}p{0.16\textwidth}
        >{\raggedright\arraybackslash}X
    }
        \toprule
        \textbf{Concept (key sources)} &
        \textbf{Object of the  loop} &
        \textbf{Channel of influence} &
        \textbf{Locus of normativity} &
        \textbf{What RSN adds} \\
        \midrule
        Model collapse \citep{shumailov2024aimodels,alemohammad2023selfconsuminggenerativemodelsmad} &
        A model's output distribution &
        Retraining on model-generated data &
        Model quality (degradation, collapse) &
        An open social loop filtered by human and institutional selection; operates without retraining \\

        Looping effects \citep{10.1093/acprof:oso/9780198524021.003.0012} & Human kinds and the people classified &Classification and self-understanding &Institutional classificatory practices &Generation of new formulations rather than classification of persons \\

        Performativity \citep{10.7551/mitpress/9780262134606.001.0001} & Markets and economic practices & Models used as guides to action & Market practices & A linguistic-semantic object; effects conditional on norm-governed uptake \\

        Recursivity and machine habitus \citep{FourcadeJohns2020,Airoldi2022} &User dispositions; outputs of classification and recommendation & Sorting, ranking, recommending & Implicit in dispositions & Specification for semantic regularities; criteria for the passage from regularity to norm \\

        AI-mediated communication \citep{10.1093/jcmc/zmz022,10.1145/3544548.3581196,PadmakumarHe2024} &Messages, attitudes, interpersonal outcomes &Assisted writing within sessions &Interpersonal evaluation &Aggregation from session-level effects to community-level semantic structure\\

        Machine culture \citep{BrinkmannEtAl2023,doi:10.1126/science.adt9819} &Cultural traits in general &Machine variation, transmission, and selection &Not specified &A domain-specific mechanism with threshold conditions, a counterfactual baseline, and a normative analysis\\

        Epistemology of AI testimony and reliance \citep{Ryan2020,Alvarado2023} &Individual beliefs formed on the basis of AI outputs &Testimony, reliance, deference &Norms of justified belief and trust &Effects on shared expressive and conceptual resources and on the conditions of vigilance and diversity\\
        
        \bottomrule
    \end{tabularx}
\end{table}

\section{Recursive semantic normalization: the mechanism}
\subsection{Terms and levels of description}

For terminological clarity, we distinguish \emph{semantic regularity} from \emph{semantic norm}. A semantic regularity is a recurrent pattern in linguistic use or interpretation: the expressions recurrently used for a content, the collocations and framings in which it appears, the distinctions drawn, and the inferences treated as licensed. A semantic norm involves a further, normative dimension in virtue of which a pattern is treated as appropriate, intelligible, admissible, or subject to correction; Section 5 states criteria for this further dimension. \emph{Normalization} refers to the process through which some regularities become increasingly stable, salient, or default within a practice; it does not entail that every recurring pattern is already a norm.

We borrow and broaden Austin's term \emph{uptake} \citep{Austin1962}. For Austin, an illocutionary act is not successfully performed unless the hearer understands its meaning and force. Here, uptake denotes the socially consequential incorporation of a formulation: its being interpreted, selected, modified, reused, amplified, or institutionally retained. Uptake includes \emph{appropriation}, in which a formulation first encountered in machine output reappears in language that a human subsequently produces unassisted. Three further notions must be kept apart. \emph{Recurrence} means that a formulation appears or is used repeatedly. \emph{Feedback} means that an earlier output affects a later act of production, selection, or interpretation. \emph{Recursion}, in the present sense, occurs when the outcome of one cycle of a sociotechnical process becomes part of the conditions governing a subsequent cycle of the same process. RSN requires more than exposure or repetition; it requires recursion in this sense.

Semantic regularities can be described at several levels: \emph{lexical} (the choice of words), \emph{phraseological} and \emph{generic} (collocations, templates, and the move structure of genres), \emph{framing} and \emph{argumentative} (the perspectives, stances, and argument schemata in which contents are presented), and \emph{conceptual} or \emph{inferential} (the distinctions drawn and the inferences treated as licensed). RSN can occur at any of these levels, but its semantic significance increases with depth while the ease of obtaining evidence decreases with it. This asymmetry structures the research programme proposed in Section 7.

\subsection{From normative practice to statistical generation and back}

The recursive relation is not adequately described as a passage from human language to machine statistics and back. Linguistic forms do not enter computational systems as normatively neutral data. The textual practices on which generative models are trained already embody historically sedimented distinctions concerning what counts as appropriate, intelligible, conventional, salient, or otherwise admissible; the typifications of everyday life, in Berger and Luckmann's  terms \citep{berger1966social}, are sedimented and transmitted above all in language. Model training therefore constitutes a \emph{statistical registration of a normatively structured linguistic environment}, not an acquisition of normativity. The model then generates new formulations from within this statistically structured space, and these formulations re-enter practices in which linguistic forms are once again evaluated according to socially and institutionally operative norms.

The resulting relation is neither a conversion of norms into statistics nor a derivation of norms from statistical frequency. It is a recursive coupling between two differently constituted orders. A normatively organized linguistic practice is statistically encoded and computationally regenerated; the regenerated forms can then be selected, rejected, appropriated, and stabilized within norm-governed practices; and, where such uptake is sufficiently recurrent, the resulting forms alter the distribution of expressions, distinctions, and inferential trajectories available to subsequent actors. We call this a \emph{normative-statistical recursion}: linguistic normativity structures the material from which statistical regularities are learned; statistical generation produces new linguistic candidates; and norm-governed uptake feeds some of those candidates back into the evolving linguistic environment from which later linguistic production, and potentially later computational systems, are constituted.

Social theory offers a precise name for the general form of this relation. For Giddens, the structural properties of social systems are both the medium and the outcome of the practices they recursively organize \citep{giddens1984constitution}. Semantic regularities are structural in this sense: speakers draw on them in order to communicate and thereby reproduce or modify them. What generative AI changes is that part of the medium is now \emph{regenerated by a computational system from the outcome of earlier practice} and offered back at the point of articulation. Normative force continues to arise through uptake rather than from the machine output itself. Human agents still evaluate, correct, select, endorse, reject, and reuse formulations, and these practices remain central to the stabilization of linguistic norms \citep{Wittgenstein1953,10.1093/0199246297.001.0001,Brandom1994}. Under computational mediation, however, these activities increasingly take place within a linguistic field that includes machine-generated alternatives. Human agents are therefore not only producers of linguistic material but also selectors and validators of computationally proposed formulations.

\subsection{The cycle}

The mechanism can be (crudely) stated as variations of a simple sequence:

\begin{center}
\begin{tikzpicture}[
  node distance=0.35cm,
  stage/.style={
    draw,
    rounded corners=3pt,
    align=center,
    minimum height=0.9cm,
    text width=2.5cm,
    font=\small,
    inner sep=5pt
  },
  arrow/.style={-{Stealth[length=2mm]}, thick}
]

\node[stage] (a) {Machine\\generation};
\node[stage, right=of a] (b) {Social\\uptake};
\node[stage, right=of b] (c) {Selection\\and evaluation};
\node[stage, right=of c] (d) {Persistence};
\node[stage, right=of d, text width=3cm] (e) {Stabilization or\\transformation of the\\linguistic environment};

\draw[arrow] (a) -- (b);
\draw[arrow] (b) -- (c);
\draw[arrow] (c) -- (d);
\draw[arrow] (d) -- (e);

\end{tikzpicture}
\end{center}

Two orders of recursion should be distinguished (Fig. 1). \emph{First-order}, or social, recursion occurs when the altered linguistic environment conditions later cycles of human and institutional production and interpretation: what has been taken up and retained becomes part of what speakers encounter, expect, and draw on. \emph{Second-order}, or computational, recursion occurs when the altered environment also enters the corpora on which later models are trained or adapted, so that subsequent generation is itself conditioned by earlier uptake. Second-order recursion intensifies the process and links it to the model-collapse literature, but it is not required for RSN: what matters is that the outcome of one cycle becomes relevant to the conditions of a later one.

\begin{figure}[t]
	\centering
    \includegraphics[width=.65\textwidth]{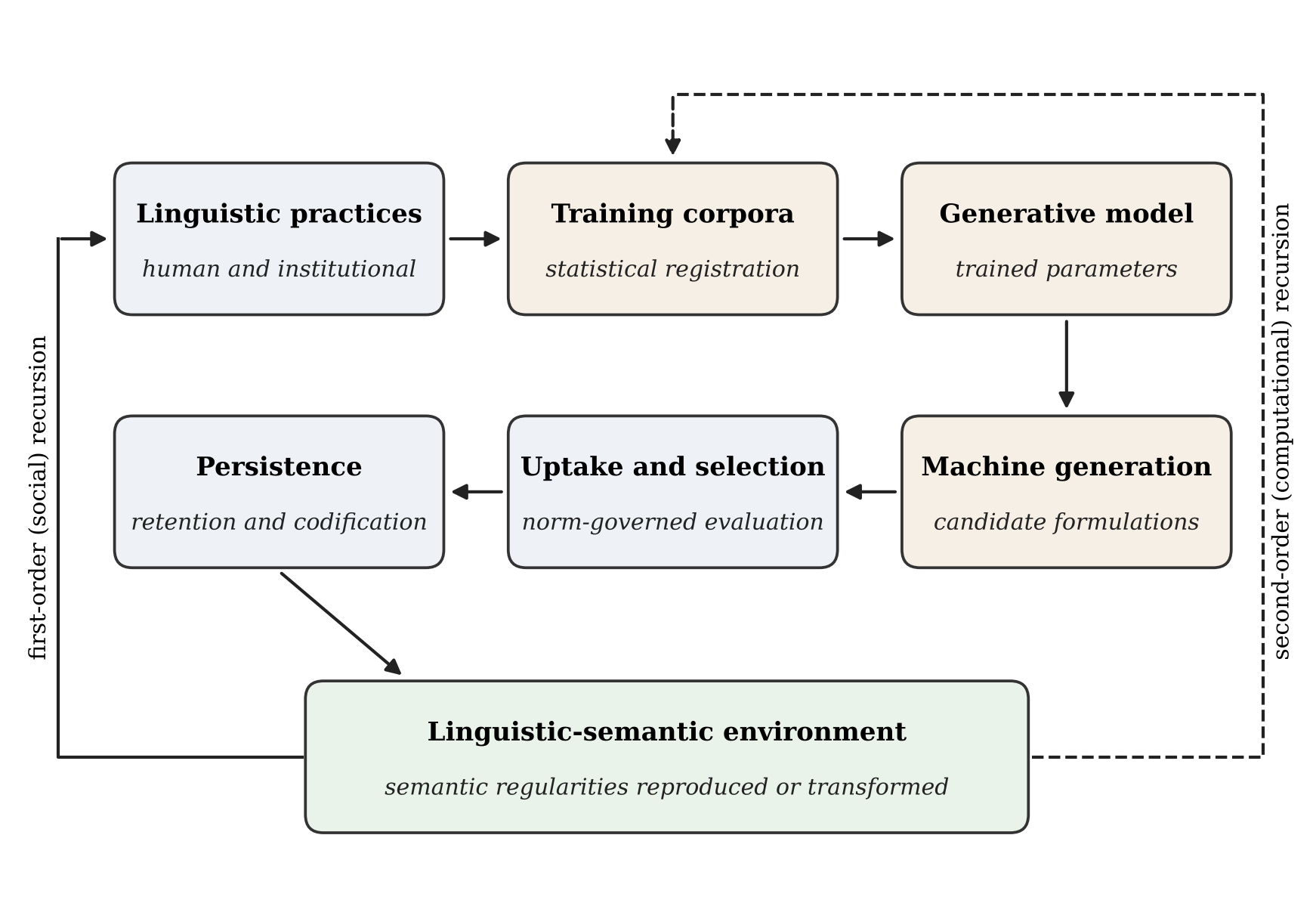}
	\caption{The hybrid semantic system and its two orders of recursion. Solid return arrow: first-order (social) recursion, through which the altered environment conditions later practice. Dashed return arrow: second-order (computational) recursion, through which it also enters the corpora of later models. Components correspond to the analytical model in Sect. 3.4}
	\label{fig1}
\end{figure}

The sequence is neither deterministic nor necessarily linear. Its analytical purpose is to distinguish exposure from incorporation, repetition from stabilization, and frequency from normative significance. A generated formulation may be ignored, rejected, or remain socially inert; a formulation can be frequent without becoming a norm, just as it can acquire institutional importance without becoming numerically dominant.

\subsection{An analytical model of the mechanism}

The sequence can be developed into an analytical model that specifies what must be present for RSN to occur and what empirical work would have to establish. The model is analogous to the way cultural-evolutionary theory describes biased transmission, in which some cultural variants are more likely than others to be adopted and passed on \citep{BoydRicherson1985}. It has six components.

The first is a \emph{domain of practice}: a bounded setting in which a recurrent communicative task is performed, such as writing abstracts in a scientific field, giving written feedback in a school system, or issuing press releases in public administration. Within such a domain, a range of alternative formulations is available for the task, and these circulate with different frequencies; this distribution of formulations is what we call the domain's \emph{linguistic environment}. The second component is the \emph{candidate space}: the range of formulations that generative models actually propose in the domain, which depends on their training data, alignment, decoding, and the prompts with which they are used. The third is \emph{selection}: the norm-governed standards by which different classes of actors (individual users, organizations, professional and educational authorities) judge which machine-generated candidates are acceptable. The fourth is the \emph{uptake share}: how often, across the communicative episodes of the domain, actors take up machine-generated candidates, whether directly or by appropriation. The fifth is \emph{persistence}: how far taken-up material remains in circulation through retention, reuse, or codification rather than disappearing after a single use. The sixth is a \emph{counterfactual baseline}: the linguistic environment that the domain would have had if no machine-generated material had been taken up, which includes ordinary semantic change, cultural transmission, and institutional diffusion.

On this model, the \emph{machine-attributable} change in a domain is the difference between the linguistic environment actually observed and the counterfactual baseline. Its size depends on two things: how much machine-derived material is taken up and retained, and how far the material that actors select from machine output differs from what would otherwise have circulated. Five propositions follow, each corresponding to a claim of the verbal account.

(P1) \emph{Generation is not sufficient}. Machine generation affects the linguistic environment only through uptake and persistence. However much material is generated, exposure without uptake, or uptake without persistence, leaves the environment unchanged.

(P2) \emph{Reproduction and transformation are distinct}. Where the formulations that actors select from machine output coincide with those that would have circulated anyway, machine-mediated circulation reproduces existing regularities: the frequencies of formulations do not change, although part of their maintenance now runs through a computational channel. Transformation occurs only where the selected material differs from the baseline in a domain in which uptake and persistence are non-negligible. RSN therefore covers both the reproduction and the transformation of semantic regularities, and the two leave different traces: transformation is visible in a comparison with the counterfactual baseline, whereas reproduction is visible only in the provenance of circulating material (Sect. 7).

(P3) \emph{The power of uptake is unevenly distributed}. Actors whose uptake is highly persistent, such as institutions that codify formulations in templates, rubrics, or style guides, can make a given amount of uptake far more consequential than actors whose uptake is ephemeral. Moreover, a formulation that models do not propose cannot be selected through the machine channel at all, whatever the standards of those who select: those who shape the candidate space, through training data, alignment, decoding, and interface design, set the bounds within which uptake can amplify anything.

(P4) \emph{Shared generation correlates selection}. When many actors draw on the same few models, they select from the same candidate space. Where their standards are similar, their contributions push the linguistic environment in the same direction, and where uptake is largely uncritical, any narrowing of the candidate space is passed on to the environment in proportion to the amount of uptake. Machine-generated variation thus differs from dispersed human innovation in being correlated at its source.

(P5) \emph{Small but sustained uptake can have cumulative effects}. If, period after period, a modest share of what circulates in a domain is taken from a stable repertoire of machine-selected formulations and retained, the linguistic environment moves step by step toward that repertoire, even when the effect in any single period is too small to notice. The cumulative significance of RSN is therefore not bounded by the size of its effect at any one moment.

Two forms of feedback extend the model. Under second-order recursion, the candidate space itself comes to depend on earlier states of the linguistic environment, because later models are trained on corpora that include material retained from earlier cycles; a regularity is then \emph{normalized} in the strong sense when the coupled loop reproduces it, that is, when machine-mediated circulation no longer moves it away from where it stands. Under what we call \emph{normative feedback}, the standards by which candidates are judged come to depend on earlier states of the environment, because what has become familiar comes to seem appropriate. Normative feedback marks the passage from regularity to norm discussed in Section 5.

The model is deliberately minimal. It treats the range of alternative formulations in a domain as given, whereas generation can introduce formulations not previously in circulation; it groups actors into broad classes and abstracts from the structure of their networks; and it considers one domain at a time. These simplifications do not affect the five propositions, which follow from the structure of the model alone, but they mean that applying it empirically requires that domains, alternatives, and classes of actors can be clearly delimited.

\subsection{Thresholds and boundary conditions}

Not every instance of machine-generated language constitutes RSN. Mere generation, exposure, or one-time use is insufficient. The relevant threshold is reached when machine-generated material is recurrently taken up, evaluated, corrected, reused, or institutionally stabilized in ways that affect subsequent linguistic practice; in the terms of Sect. 3.4, when uptake and persistence are non-negligible in a domain and the selected material either alters the linguistic environment relative to its counterfactual baseline or comes to carry a significant share of its maintenance.

The framework is deliberately narrower than a general theory of AI influence. Machine-generated language that is ignored or rejected does not constitute RSN; nor does isolated use, repetition that leaves the relevant semantic environment unchanged, or a linguistic change fully explained by pre-existing processes of semantic change, cultural transmission, or institutional diffusion. The framework applies only where there is a plausible pathway from machine-mediated generation through social uptake to subsequent stabilization or transformation of the linguistic-semantic environment. This boundary prevents the concept from becoming a retrospective label for any linguistic change that happens to occur after the introduction of generative AI.

\section{From individual agents to hybrid semantic systems}

\subsection{The unit of analysis}

If RSN is the mechanism, the next question is what unit of analysis it requires. The relevant unit cannot be the isolated human subject, the computational model taken on its own, or a simple aggregation of the two. It is instead the \emph{hybrid semantic system}: a recursively coupled arrangement in which human practices, computational generation, and institutional uptake jointly shape the conditions under which linguistic forms become available, intelligible, repeatable, and stable. In the terms of Sect. 3.4, a hybrid semantic system is a domain of practice together with the actors, models, and institutions whose generation, selection, uptake, and persistence jointly shape the evolution of its linguistic environment.

This formulation does not imply that AI systems are semantic agents in the same sense as human participants. Their significance lies in their participation in processes of linguistic production, selection, and circulation. Human prompts, evaluations, corrections, and institutional uses constrain the outputs that computational systems generate; those outputs can re-enter human interpretation and influence subsequent formulations. The claim is not that meaning resides in the machine, but that, under conditions of sustained computational mediation, semantic stabilization increasingly occurs within a coupled field of human and computational activity.

The proposed unit is related to, but distinct from, several established approaches. Distributed cognition and the extended mind thesis show how cognitive processes may extend beyond the individual into artifacts and environments \citep{Hutchins1995,ClarkChalmers1998}. Actor-network theory emphasizes heterogeneous relations among human and non-human actants \citep{Latour2005}. Postphenomenology analyses how technologies mediate perception, action, and moral practice \citep{Ihde1990,Verbeek2005,Verbeek2011,Coeckelbergh2017} has argued that language and technology must be thought together rather than as separate domains. Systems theory shifts explanatory attention from individual subjects to communication \citep{Luhmann1995} and distinguishes social structure from semantics, the stock of condensed and repeatable forms of meaning that a society keeps available for communication \citep{Luhmann2012}; in these terms, RSN describes a new route by which a society's semantics is replenished. \cite{10.7551/mitpress/14189.001.0001} has argued that algorithms can participate in communication without understanding, provided that their contributions are understood by those who receive them. Externalist semantics, finally, already locates meaning partly in a division of linguistic labour \citep{Putnam1975}. These perspectives are relevant to the present argument, but they do not by themselves specify the mechanism examined here. The hybrid semantic system can be read as the condition in which the division of linguistic labour acquires a non-human participant whose contributions are statistically derived from that very division and are fed back into it. Mediation is therefore a background condition of the process; RSN is the specific mechanism the present paper seeks to isolate.

\subsection{Relational hybridity: weak and strong}

Technohybridism does not suggest that humans become machines, that machines become human subjects, or that computational systems independently generate semantic norms. Nor does it deny the continuing importance of human agency. It proposes that the processes through which semantic regularities are produced and stabilized are increasingly distributed across a configuration in which human participants, computational systems, and institutional settings recursively affect one another. The relevant hybridity is therefore relational and processual rather than ontological.

Two levels of hybridity can be distinguished. \emph{Weak hybridity} refers to the fact that computational systems are historically derived from human linguistic and cultural material: they are not ontologically foreign participants entering an otherwise purely human semantic order. \emph{Strong hybridity} refers to the condition that emerges after deployment, when such historically derived systems become recurrently coupled to human practices and institutions and thereby participate in the reproduction and transformation of the linguistic environment. The distinction marks two levels of analysis, not two kinds of entity. Weak hybridity concerns the historical constitution of the computational component; strong hybridity concerns the relational configuration that emerges when that component becomes embedded in ongoing social practice. Technohybridism therefore posits no original opposition between human and machine semantic domains. It examines how a historically human-derived technology can become part of the recursive processes through which semantic regularities are subsequently produced, selected, and stabilized.

\subsection{Semantic fields, normalization fields, and novelty}

To characterize the mechanism more precisely, it is useful to distinguish a \emph{semantic field} from a \emph{normalization field}. A semantic field is the structured space of expressions, inferential pathways, and conceptual trajectories available at a given time. A normalization field is the set of social, institutional, technological, and historical constraints that make some of these trajectories more available, salient, admissible, or stable than others. In the model of Sect. 3.4 the semantic field corresponds to the range of alternative formulations in a domain and the frequencies with which they circulate, while the normalization field corresponds to the factors that govern how these frequencies change: ordinary processes of linguistic change, the candidate space offered by models, the standards of selection, and the extent of uptake and persistence. The distinctive feature of technohybrid settings is that computationally mediated interaction becomes one of the processes through which the normalization field itself is modified.

The same distinction clarifies the problem of semantic novelty. Novelty does not arise in the absence of constraints: both human and computational linguistic activity operate within historically structured spaces of possibility. Human semantic activity is shaped by linguistic practice, institutions, education, embodiment, and historical inheritance; computational generation is shaped by training data, model architecture, optimization, alignment, and prompt context. In both cases, novel formulations can emerge through the recombination, extension, or reweighting of inherited semantic relations. Under technohybrid conditions, the relevant question is therefore how interaction between these differently constrained systems affects which semantic trajectories become available for subsequent uptake.

The coupling of human and computational systems does not simply place two sets of constraints alongside one another. Through recursive interaction, each can affect the conditions under which the other operates: prompts, selections, and evaluations influence subsequent computational outputs, while those outputs can reshape later human articulation and interpretation. \emph{Hybrid normalization} refers to this transformation of the normalization field through recurrent interaction between human and computational processes; in the terms of Sect. 3.4, it corresponds to second-order recursion and normative feedback. The concept remains subordinate to RSN: it specifies how the field of semantic stabilization can change as a result of the coupling. The resulting picture is neither one of autonomous machine agency nor one of unchanged human linguistic control. Human agents remain responsible for interpretation, evaluation, and normative uptake, while computational systems increasingly participate in determining which formulations are presented, repeated, and available for further stabilization.

\section{From regularity to norm: normativity, power, and social practice}

\subsection{When does a machine-derived regularity become a norm?}
The distinction between regularity and norm is essential, and it is old. That a pattern of use is frequent does not make it correct; the rule-following considerations show that no finite pattern of past use by itself determines what counts as going on in the same way \citep{Wittgenstein1953,Kripke1982}. On normative pragmatist accounts, norms are implicit in practice: they consist in practitioners' treating performances as correct or incorrect, attributing and acknowledging commitments and entitlements, and sanctioning deviation \citep{Brandom1994}. Accounts of convention and social norms specify the expectations that sustain such practices. A convention is a regularity sustained by mutual expectations of conformity \citep{Lewis1969}; a social norm, on Bicchieri's account \citep{Bicchieri2006}, obtains when people prefer to conform to a rule on condition that they expect others to conform (empirical expectations) and believe that others expect them to conform and may sanction deviation (normative expectations).

These accounts yield three markers by which a machine-derived semantic regularity can be said to have acquired normative standing in a practice.

(N1) \emph{Correction}. Deviations from the machine-derived form are treated as errors and corrected toward it, by editors, reviewers, teachers, assessors, or automated checking procedures.

(N2) \emph{Expectation}. Participants expect others to use the form and believe that others expect them to use it, so that its absence becomes noticeable and calls for justification.

(N3) \emph{Codification}. The form enters templates, style guides, rubrics, model answers, administrative boilerplate, or training materials, where it is made explicit and binding for subsequent participants.

None of these markers requires that participants know or believe the form to be machine-derived. In the terms of Sect. 3.4, they correspond to normative feedback: the standards by which candidates are judged come to favour formulations because these have become familiar and expected, and codification makes their persistence close to complete. Consider an illustrative case from education. Suppose that model-generated phrasings for assessment criteria or written feedback are adopted by teachers, reused across classes, and incorporated into rubrics against which students' work is evaluated. At that point the phrasings are no longer only frequent: student writing that departs from them may be marked down (N1), teachers and students come to expect them (N2), and they are written into the instruments of assessment (N3). The formulation has passed from statistical candidate to semantic norm, and it has done so entirely through human and institutional uptake. Analogous trajectories can be described for reviewers' expectations about the phrasing of scientific abstracts or for the boilerplate of administrative communication. These are illustrations of the conditions under which the passage would occur, not claims that it has occurred in any particular setting; whether it has is an empirical matter (Sect. 7).

\subsection{The power of uptake}

Uptake should not be treated as a neutral or homogeneous human operation. Bourdieu argued that the value of linguistic products depends on the market in which they are exchanged and on the authority of those who produce and sanction them, and that the educational system plays a decisive part in establishing and legitimating an official language \citep{bourdieu1991language} . RSN inherits this asymmetry. Different actors and institutions control different stages of selection, amplification, and institutionalization, and the representation of Sect. 3.4 makes these stages explicit.

Developers and interface designers shape the candidate space before any user selects from it. Likelihood-maximizing decoding tends to produce bland and repetitive text \citep{HoltzmanEtAl2020}, reinforcement learning from human feedback reduces the diversity of model outputs \citep{ICLR2024_5a68d050}, and the opinions reflected in model outputs are unevenly aligned with those of different demographic groups \citep{pmlr-v202-santurkar23a}. Such choices define the candidate space and, by proposition (P3), bound what uptake can amplify. Users select, reject, or modify outputs, thereby setting the standards of selection and the extent of uptake. Organizations determine which formulations enter workflows, and professional, educational, and regulatory authorities determine which forms acquire legitimacy and durability, thereby governing persistence. Because a small number of models serve very large populations, many actors select from the same candidate space, and research on algorithmic monoculture shows how such sharing can correlate outcomes across otherwise independent decision-makers \citep{doi:10.1073/pnas.2018340118,3600270.3600535}. By proposition (P4), the same holds for semantic proposals.

RSN is therefore compatible with continuing human normative agency while recognizing that the capacity to make uptake consequential is unevenly distributed. The relevant question is not only whether users adopt machine-generated language, but who can make such adoption consequential.

\subsection{Uptake as social practice: routines, isomorphism, and stratification}

Uptake has so far been described as selection among candidates. Sociologically, however, most uptake is not a deliberate choice but a component of practice. Practice theory understands social life as composed of routinized bundles of doings and sayings, held together by shared understandings, rules, and ends \citep{Schatzki2002,Reckwitz2002}. Machine-generated formulations enter such routines when drafting with a model becomes the normal way of producing a type of text, and they are then reproduced less through evaluation than through habit. The conditions of practice matter accordingly. In peer reviews submitted to AI conferences, the estimated share of LLM-modified text was higher in reviews submitted close to the deadline and in reviews whose authors reported lower confidence \citep{pmlr-v235-liang24b}: uptake intensifies where time is scarce and certainty is low. The intensity of RSN in a domain is therefore likely to depend on its organization of time and workload as much as on the attitudes of individuals.

Institutional diffusion follows familiar patterns. \cite{DiMaggioPowell1983} distinguished three mechanisms through which organizations come to resemble one another: coercive isomorphism, through rules and dependence; mimetic isomorphism, through imitation under uncertainty; and normative isomorphism, through professionalization. Each has a semantic counterpart. Machine-derived formulations can be imposed through mandated templates and assessment instruments, copied from organizations perceived as successful when the appropriate wording of a new kind of document is uncertain, and diffused through professional training and networks. Because organizations increasingly draw on the same few models, these mechanisms may operate on formulations that are already convergent at their source, so that institutional isomorphism and the correlation of machine proposals (P4) reinforce one another.

Uptake is also socially stratified. Adoption of LLM-assisted writing differs across social contexts. In consumer complaints submitted to a US federal agency, estimated adoption stabilized at higher levels in highly urbanized than in other areas and, more modestly, at higher levels in areas with lower educational attainment; in job postings, it was higher among smaller and younger firms \citep{LIANG2025101366}. Bourdieu's 1991 analysis of linguistic capital suggests an interpretation \citep{bourdieu1991language}. Where command of the legitimate language of complaints, applications, or official correspondence is unequally distributed, language models can function as a technology of access, supplying formulations that carry institutional legitimacy to those who lack them. Such compensatory use is ambivalent. It may reduce one form of linguistic inequality, but it does so by aligning the expression of less advantaged speakers with formulations that models make standard, so that their semantic repertoire becomes more tightly coupled to the hybrid system than that of speakers who can afford to write without it. Conversely, distinction can work against RSN \cite{Bourdieu1984}. Once certain formulations are publicly identified as markers of machine-generated text, avoiding them can become a sign of competence. The marked decline of the word "delve" in arXiv abstracts after it was publicized as characteristic of ChatGPT, which \cite{geng-trotta-2025-human} interpret as authors adapting how they select and edit model output, is consistent with such reflexive avoidance. RSN is therefore not a uniform process: its direction and intensity depend on the positions, resources, and strategies of actors in particular fields.

\subsection{Semantic-normative co-production}

The semantic and normative dimensions should be distinguished without being separated. Computational systems generate linguistic candidates within statistically structured spaces, while human and institutional practices determine which candidates acquire practical significance. The resulting interaction can create a feedback relation in which formulations that become recurrently acceptable are more likely to be reused, while recurrent reuse in turn contributes to their stabilization. The process is not a derivation of norms from statistics; it is a process through which statistically generated linguistic material enters norm-governed practices and can thereby affect future judgments of appropriateness and inferential acceptability.

This also complicates the distinction between generation and validation. When machine-assisted formulations are repeatedly adopted, adapted, and reused, no single moment need determine their eventual semantic or normative status. The relevant outcome can instead emerge through successive acts of generation, evaluation, correction, and stabilization across the hybrid semantic system. Authority is not dissolved, but its practical location becomes distributed across the process through which formulations acquire durable social uptake.

\section{Selected consequences of recursive semantic normalization}

The consequences of RSN should be distinguished rather than conflated. \emph{Semantic stabilization} concerns the conditions under which linguistic forms become publicly available, repeatable, and socially usable; \emph{normative stabilization} concerns the standards by which such forms come to be treated as appropriate, intelligible, or inferentially acceptable; and \emph{epistemic evaluation} concerns the resources through which claims are assessed as justified, warranted, or true. RSN connects these dimensions because the same process of recurrent uptake can affect all three, while leaving their differences intact. Normative stabilization was treated in Section 5; this section addresses semantic, social-epistemic, and historical consequences. The consequences discussed below are indicative rather than exhaustive: they are those that follow most directly from the mechanism specified in Sections 3-5, and further consequences may become visible as the mechanism is studied in particular domains.

\subsection{Semantic consequences: articulation, recognition, and retrospective stabilization}

At the semantic level, RSN affects the conditions under which contents become publicly articulable, recognizable, and socially usable. The argument does not require treating meaning as a private mental object that is simply expressed outwardly. It emphasizes instead the role of recurrent uptake, correction, recognition, and reuse in stabilizing publicly available formulations, consistent with accounts on which meaning in conversation is established through collaborative grounding \citep{Clark1996}. In hybrid settings, machine-generated language can contribute to the selection among alternative ways of articulating a content that is initially only partially determinate.

The consequence is especially visible where a speaker has an antecedent intention or inchoate content without a fully articulated expression of it. A scientist may begin with an intuition that is not yet fully articulable; an LLM generates several possible formulations, and one of them "clicks": the scientist judges that it captures what was meant. The model does not thereby become the possessor of the meaning, and the scientist does not simply retrieve a pre-formed content from the machine. Rather, the interaction contributes to settling one semantic trajectory among several possibilities by providing a formulation that is subsequently recognized, adopted, and developed. We call this \emph{retrospective stabilization}: once a formulation is recognized as apt, it can become the preferred articulation through which the speaker subsequently understands, communicates, and develops the original intuition.

The claim should remain limited. It does not follow that all meaning is interactionally constituted, even if meaning is understood in relational rather than exclusively intrinsic terms, nor that machine mediation is necessary for semantic determination. Under ordinary conditions, articulation already plays a role in determining how an initially underdetermined thought becomes publicly available. Under technohybrid conditions, this process can involve a computationally generated set of candidate formulations, and the distinction between generation and validation becomes less sharp: the machine supplies candidates, while human recognition and subsequent uptake determine whether any of them acquires durable significance. Psycholinguistic research suggests that such uptake need not be deliberate. Interlocutors align their lexical and syntactic choices with one another in dialogue \citep{PickeringGarrod2004}, and they also align with computer interlocutors \citep{BraniganEtAl2010}; the adoption of chatbot word choices after brief exposure \citep{yakura2026empiricalevidencelargelanguage} is consistent with this mechanism.

At the social level, the possibility matters because such interactions are not confined to isolated individuals. Similar processes occur across scientific, legal, educational, artistic, and political settings, and, by proposition (P4), the candidates offered to very many speakers are drawn from the same few models. Retrospective stabilization thus scales from an individual episode to a correlated population process. The semantic consequence of RSN is not the replacement of human meaning by machine output, but a modification of the pathways through which contents become expressible, shareable, and publicly legible.

\subsection{Social-epistemic consequences}

At the epistemic level, RSN affects the background resources through which claims are formulated, assessed, justified, and rejected. The point is not that truth becomes inaccessible or that truth-conditions cease to matter. It is that the conceptual and inferential resources through which truth-directed evaluation takes place may themselves be partly shaped by the recursive linguistic environment in which claims are produced and circulated. \cite{habermas1985theory} describes the lifeworld as a background reservoir of taken-for-granted interpretive patterns on which participants draw in reaching understanding; RSN identifies one route by which that background can be modified without ever becoming an explicit theme of communication.

If machine-mediated discourse repeatedly supplies formulations, distinctions, and inferential pathways that are subsequently adopted by users and institutions, then the background of epistemic evaluation may no longer be entirely independent of the processes that shape it. The epistemic question is then not only whether a particular claim is true or justified, but also how the conceptual repertoire within which it is formulated became established, which alternatives became less salient, and how machine-mediated circulation may have influenced the background from which subsequent questions and judgments arise. \cite{MesseriCrockett2024} warn that reliance on AI in research can produce "monocultures of knowing", in which some methods and perspectives crowd out others; \cite{doi:10.1126/science.adi0248} argue that the confident and fluent presentation of generative outputs can distort human beliefs. RSN locates a further and less visible pathway: the shaping not of particular beliefs but of the repertoire in which beliefs are formulated. Four social-epistemic dimensions of this pathway can be distinguished.

The first concerns epistemic vigilance. Communities protect themselves against error and manipulation partly by calibrating trust to sources \citep{SperberEtAl2010}. Appropriation (Sect. 3.1) severs machine-originated formulations from their source: once a formulation has been adopted into a speaker's own writing or speech, hearers assess it as that speaker's assertion, with the vigilance appropriate to that speaker, and its machine origin is no longer visible to hearers and, at times, not even to the speaker. We call this \emph{provenance laundering}. It does not make claims false, but it means that semantic defaults supplied by a small number of models circulate under the credibility of many dispersed human speakers, beyond the reach of the source-sensitive scrutiny that social epistemology identifies as a safeguard of testimonial knowledge. Disclosure rules that attach to documents rather than to formulations cannot fully address this, because appropriated formulations outlive the documents in which they first appeared.

The second concerns cognitive diversity. Accounts of social objectivity hold that a community's knowledge is reliable to the extent that it contains diverse perspectives capable of transformative criticism \citep{Longino1990}, and models of scientific communities show that a transient diversity of approaches protects against premature convergence on inferior views \citep{Zollman2010,Kitcher1990,HongPage2004}. Homogenization at the level of framings and conceptual distinctions, the upper rungs of the ladder in Sect. 7, would therefore matter epistemically even if every homogenized formulation were accurate: it would reduce the range of perspectives from which accepted views can be criticized. Because machine proposals are correlated at their source (P4), RSN can produce such convergence without any coordination among those who take them up.

The third concerns hermeneutical resources, and it has a distributive dimension. \cite{Fricker2007} describes hermeneutical injustice as a situation in which a gap in collective interpretive resources puts some people at a disadvantage in making sense of their own experience. Where ways of articulating experience that are characteristic of marginalized groups or of smaller language communities are under-represented in training data, as web-scale corpora tend to over-represent hegemonic viewpoints \citep{BenderEtAl2021}, and are therefore rarely proposed by models, RSN can narrow collective hermeneutical resources even if every individual act of uptake is voluntary, a risk that the documented drift of assisted writing toward Western styles makes concrete \citep{10.1145/3706598.3713564}.

The fourth concerns the form of trust involved. \cite{Luhmann1988} distinguished trust, which involves a consciously accepted risk, from confidence, the taken-for-granted expectation that things will go on as usual, and \cite{Giddens1990} described how modern actors depend on abstract systems whose workings they cannot assess. Routinized uptake of machine-generated formulations (Sect. 5.3) is better described as confidence than as trust: formulations are not judged reliable after consideration but are accepted because they arrive in the familiar flow of work. This matters because the safeguards of social epistemology, vigilance, criticism, and the comparison of sources, are exercised in the mode of trust rather than of confidence. Where uptake takes the form of confidence, the hybrid system reproduces its semantic defaults below the threshold at which they would become objects of epistemic evaluation.

Taken together, these dimensions connect individual acts of reliance to collective epistemic conditions. The relevant consequence is therefore one of epistemic reflexivity rather than scepticism. Users and institutions may need to attend not only to the correctness of particular outputs, but also to the formation of the semantic and inferential defaults through which outputs become intelligible and assessable, and to the social arrangements, such as disclosure, diversity of tools, and occasions for deliberate rather than routine use, that keep those defaults open to criticism.

\subsection{Historical consequences: classical and technohybrid normalization}

RSN also has a historical dimension. AI systems are trained on human-produced language, which already embodies historically accumulated meanings, distinctions, norms, and inferential patterns. In this sense, generative systems are initially constituted within a pre-existing semantic order. Once deployed, however, their outputs can enter the same social processes through which linguistic regularities are repeated, revised, and extended. The source of semantic normalization consequently becomes more distributed: human practices remain constitutive, but computationally generated language can increasingly contribute to the material that those practices evaluate, reuse, and stabilize.

This does not amount to an absolute break with earlier media. Writing, print, dictionaries, standardized schooling, and search technologies already transformed the circulation of language and the stabilization of meaning \citep{bourdieu1991language,doi:10.1126/science.adt9819}. The historical novelty of RSN lies in the structure of a feedback relation rather than in mediation as such, and it can be specified by five features whose combination is new. Generative systems generate context-sensitive candidates rather than retrieving or reproducing existing texts; they do so at the point of articulation, in response to a speaker's own half-formed intention; they operate at a scale and speed that allow rapid uptake across many domains at once; their proposals are centralized, since a small number of models, developed by a few firms that increasingly dominate the computing power, data, and expertise on which such models depend \citep{AhmedWahedThompson2023}, offer candidates to very large populations, so that variation is correlated at its source (P4); and the selection that shaped the candidate space, through data curation, alignment, and decoding, is largely opaque to those who take up its products. Print standardized language through identifiable codifiers, institutions, and reference works; technohybrid normalization can approximate the homogenizing effect of a standard without an explicit codifier. The historical claim is consequently one of structural difference rather than absolute discontinuity: earlier media transformed linguistic circulation, whereas generative AI adds a direct and recurrent form of machine-mediated linguistic production to that circulation.

\section{What would count as evidence: an empirical programme}

\subsection{An evidential ladder}

Although the present paper is conceptual, the framework yields specific empirical expectations. RSN should be more likely where machine-generated formulations are repeatedly taken up, remain persistent across communicative episodes, become embedded in institutional routines, or substantially affect the range or salience of linguistic alternatives. Conversely, frequent exposure without persistence, or persistent change without a plausible pathway through machine-generated material, would count against the explanation. Table 2 organizes the expectations by the levels of description distinguished in Sect. 3.1.

\begin{table}[t]
    \caption{An evidential ladder for recursive semantic normalization. Rows 1-4, the rungs of the ladder, correspond to the levels of description in Sect. 3.1; semantic significance increases, and currently available evidence decreases, from row 1 to row 4. Row 5 lists the normative markers of Sect. 5.1, which cut across rows 1-4}
    \label{tab:evidlad}
    \centering
    \small
    \setlength{\tabcolsep}{3pt}
    \renewcommand{\arraystretch}{1.2}
    \begin{tabularx}{\textwidth}{
        >{\raggedright\arraybackslash}p{0.20\textwidth}
        >{\raggedright\arraybackslash}p{0.15\textwidth}
        >{\raggedright\arraybackslash}p{0.17\textwidth}
        >{\raggedright\arraybackslash}p{0.16\textwidth}
        >{\raggedright\arraybackslash}X
    }
        \toprule
        \textbf{Rung and unit} &
        \textbf{Example indicators} &
        \textbf{Research designs} &
        \textbf{Current evidence} &
        \textbf{Pattern counting against RSN} \\
        \midrule
        1 Lexical: words and their frequencies&Frequency of words characteristic of model output relative to counterfactual trends; adoption in unassisted speech&Excess-vocabulary analysis; distributional estimation; synthetic control; exposure experiments&Substantial for academic, institutional, and spoken English \citep{doi:10.1126/sciadv.adt3813,pmlr-v235-liang24b,LIANG2025101366,yakura2026empiricalevidencelargelanguage}&No deviation from pre-adoption trends, or equal deviation in low-adoption domains \\

        2 Phraseological and generic: collocations, templates, move structure of genres&Convergence of n-gram and move-structure profiles; declining between-author variance within genres&Corpus stylometry against counterfactual baselines; difference-in-differences across adoption levels&Emerging \citep{10.1145/3706598.3713564,PadmakumarHe2024}& Convergence confined to texts with machine provenance and absent from later unassisted texts\\

        3 Framing and argumentative: perspectives, stances, argument schemata & Shifts in the distribution of framings or argument types; reduced diversity of ideas across writers & Controlled co-writing experiments; panel analysis of public and institutional discourse & Experimental and session-bound \citep{10.1145/3544548.3581196,doi:10.1126/sciadv.adn5290,10.1145/3635636.3656204}  & Effects that dissipate after sessions and do not reappear in unassisted production \\

        4 Conceptual and inferential: distinctions, definitions, licensed inferences & Changes in definitions and operationalizations in institutional documents; diachronic shifts in the distributional semantics of key terms & Diachronic embedding analysis \citep{hamilton-etal-2016-cultural}; document panels with dated codification & None yet; hypothesis & Stable conceptual distinctions despite high uptake at lower rungs \\

        5 Normative markers N1-N3: correction, expectation, codification & Entry of machine-derived formulations into rubrics, style guides, and templates; corrections toward them & Document analysis; editorial and assessment records; surveys of empirical and normative expectations & No systematic evidence yet & Frequency without correction, expectation, or codification\\

        \bottomrule
    \end{tabularx}
\end{table}

Existing evidence is concentrated at the first rung. Excess-vocabulary analyses compare observed word frequencies with counterfactual frequencies extrapolated from pre-LLM trends \citep{doi:10.1126/sciadv.adt3813}; distributional methods estimate the share of LLM-modified text in large corpora \citep{pmlr-v235-liang24b,LIANG2025101366}; and synthetic-control analyses of spoken language combined with exposure experiments trace the appropriation of characteristic words into unassisted speech \citep{yakura2026empiricalevidencelargelanguage}. These studies indicate that two components of the mechanism, uptake and the persistence that publication or habitual use confers, are present at scale in some domains. They do not establish that semantic regularities in the stronger sense have been transformed, and a shift in the frequency of particular words is not, by itself, a change in meaning. The framework's distinctive claims concern the higher rungs, for which the evidence is at present experimental and session-bound (rung 3) or absent (rung 4).

\subsection{Identification}

The analytical model of Sect. 3.4 identifies what must be established. Machine-attributable change can be assessed only against an estimate of the counterfactual baseline. Feasible designs include extrapolation of pre-adoption trends, as in excess-vocabulary analysis; synthetic control, in which the counterfactual for a treated domain is constructed from a weighted combination of comparison domains \citep{AbadieDiamondHainmueller2010}; and difference-in-differences designs that compare domains, institutions, or language communities with different rates or timings of adoption, for instance where institutional policies restrict or mandate the use of generative tools. Uptake and persistence require evidence of provenance: distributional estimation of the share of machine-modified text, workflow logs, and longitudinal tracing of particular formulations across documents. Appropriation, which is invisible to detectors of machine-generated text, is visible only through counterfactual comparison or exposure experiments. Standards of selection can be studied experimentally, by asking participants to choose among machine-generated candidates under controlled conditions, and observationally, through editorial and assessment records. Normative markers (N1-N3) require document analysis of style guides, rubrics, editorial policies, reviewer reports, and administrative templates, ideally in panel designs that date the entry of a formulation into codified instruments relative to its first appearance in machine output. Because uptake is routinized and stratified (Sect. 5.3), these designs should be complemented by ethnographic and interview studies of writing practices in organizations, by comparisons of adoption across social positions, and by network analyses that trace the diffusion of formulations across organizations and professional communities.

Proposition (P2) implies an important caution. Where machine-mediated circulation reproduces existing regularities, counterfactual comparison will show no effect even though a growing share of their maintenance runs through the computational channel. Evidence of reproduction therefore requires provenance data and cannot be inferred from the absence of change.

\subsection{Disconfirmation}

The framework would be disconfirmed, or its scope restricted, by four patterns. First, if machine-attributable change were confined to the lexical rung despite high uptake, RSN would be real but semantically shallow. Second, if changes attributed to machine mediation appeared equally in domains with low and high adoption, or preceded adoption, the pathway condition would fail. Third, if machine-derived formulations did not persist beyond the episodes in which they were generated, the threshold condition would fail. Fourth, if machine-derived regularities became frequent without acquiring any of the normative markers, RSN would yield regularity without normativity. These indicators should be understood as empirical criteria for testing the framework rather than as findings established by the present paper. The argument is accordingly conditional: it specifies a mechanism and the observable patterns that would support or weaken it, while leaving the empirical magnitude and prevalence of RSN open to investigation.

\section{Objections and replies}

The framework describes a sociotechnical configuration rather than proposing a metaphysical redefinition of meaning. It faces objections concerning the relation between regularity and normativity, the role of human agency, the novelty of generative AI, its relation to existing accounts of recursion and to the epistemology of testimony, the significance of the evidence, and its testability. The replies delimit the framework's claims rather than multiplying qualifications.

\subsection{Statistical regularity is not normativity}

A first objection is that the framework derives normativity from statistical regularity. Language models are trained on distributions of usage, but distributions are not norms. The objection is correct, and the framework does not require such a reduction. RSN does not claim that statistical structure is itself normative in the human sense. Its claim is that statistically organized outputs can enter practices already governed by norms of interpretation, evaluation, correction, and acceptance, and can thereby become relevant to the reproduction or modification of those norms \citep{Wittgenstein1953,Brandom1994}. The distinction is between the source of a linguistic candidate and the process through which that candidate acquires normative significance. A model does not authorize a norm merely by generating a formulation; normative significance arises through uptake, evaluation, correction, and recurrent use, and Section 5 states the markers by which it can be recognized. This does not mean, however, that the material from which models learn is normatively neutral. Language use, and the formal regularities it leaves in texts, already embodies normativity, and this normativity is carried over, as a statistical trace, into what models learn (Sect. 3.2). Model outputs therefore re-enter normative practice as candidates already shaped by earlier norms, but without the authority to fix them.

\subsection{Human validation remains central}

A second objection is that the framework understates human agency. If human users select, validate, edit, and authorize machine-generated formulations, why not conclude that semantic norms remain entirely human? The framework accepts the premise but not the conclusion. Human agency remains indispensable to normative uptake, but the linguistic field within which human judgments are made can change. Under computational mediation, validation increasingly occurs among alternatives that have themselves been generated computationally, from a candidate space shaped by choices that the validators did not make (Sect. 5.2). Human agents remain evaluators and normative participants, but they do not necessarily operate within a linguistic field produced exclusively by prior human practice. The locus of normative authority should therefore be understood as distributed without being dissolved: humans remain capable of rejecting, revising, and overriding machine-generated formulations, while the structure of the field over which those judgments operate is altered. This is why the framework does not require autonomous machine normativity: its claim concerns the transformation of the conditions of human normative practice. Moreover, the standards by which humans evaluate, select, and correct formulations are themselves mediated by the semantic framework available to them at a given time, and, as argued in Sects. 3-4, that framework may already be technohybrid. Human validation is therefore not a checkpoint outside the loop but one of its moments.

\subsection{Nothing new: print, dictionaries, and search}

A third objection is historical. Writing, print, dictionaries, search engines, and other media have long shaped the circulation of language and the stabilization of meaning, so generative AI may appear to be simply another technology of linguistic mediation. The framework accepts this continuity and locates its claim more narrowly. RSN is not presented as the first historical instance in which technology affects linguistic practice. The distinctive feature lies in the combination identified in Sect. 6.3: generation of new candidates rather than retrieval, at the point of articulation, at scale, from centralized sources, through an opaque prior selection. Earlier media altered what linguistic material was available, how it circulated, and how it was retrieved; generative systems add the direct production of context-sensitive candidates that can be selected, adapted, and incorporated into later linguistic environments.

\subsection{This is already captured by looping effects, performativity, or machine habitus}

A fourth objection holds that RSN merely relabels established accounts of recursion in the sociology of algorithms. The concepts are indeed kin, and the framework draws on them, but they differ in object, channel, and normative locus (Sect. 2.4, Table 1). Looping effects concern classifications of people and the people classified \citep{10.1093/acprof:oso/9780198524021.003.0012}; performativity concerns models used as guides to economic action \citep{10.7551/mitpress/9780262134606.001.0001}; machine habitus and the recursivity of social and machine learning concern dispositions reproduced through classification, ranking, and recommendation \citep{FourcadeJohns2020,Airoldi2022}. None of them specifies how the generation of new linguistic candidates, filtered by norm-governed uptake, alters the semantic regularities of a community, or when the resulting regularities become norms. RSN can be read as the semantic specification of the broader recursivity these accounts describe.

\subsection{This is model collapse in sociological dress}

A fifth objection is that RSN restates the model-collapse problem. The two differ in loop, object, and evaluative standard (Sect. 2.2). Model collapse is a closed technical loop in which outputs are fed back as training data, and its problem is the degradation of a model. RSN is an open social loop in which outputs are filtered by human and institutional selection, and its problem is the transformation of a community's semantic environment. The two are linked by second-order recursion, and RSN identifies one route by which the technical problem arises in practice: what is socially selected and retained determines the composition of future corpora. But RSN can occur without retraining, and its consequences, for example for normative standards or hermeneutical resources, are not measured by model quality.

\subsection{This is a question for the epistemology of testimony}

A further objection holds that the relevant issues are already covered by the epistemology of testimony and of reliance on AI: the question is whether people are justified in accepting AI outputs, and existing accounts address it \citep{Ryan2020,Alvarado2023} The framework agrees that these questions are important but argues that they do not exhaust the social-epistemic problem. Testimony concerns the transmission of particular contents from identifiable sources; RSN concerns the transformation of shared expressive and conceptual resources whose machine provenance is typically lost through appropriation (Sect. 6.2). A community can be justified in each act of reliance and still find that the repertoire in which its beliefs are formulated has narrowed. The framework thus complements the epistemology of testimony with an account of the epistemic environment in which testimony takes place.

\subsection{Lexical fashions are semantically trivial}

Another objection is that the documented effects, such as the rise of particular words in scientific abstracts or podcasts, are stylistic fashions without semantic significance. The framework agrees that a shift in word frequency is not by itself a change in meaning, and it builds this point into the evidential ladder (Table 2). Lexical shifts matter as tracers: they show that uptake and persistence, two of the mechanism's components, operate at scale, and they provide measurable markers whose diffusion can be followed across domains. Whether the mechanism reaches the framing and conceptual levels, where semantic significance lies, is precisely the question the framework poses and leaves open.

\subsection{The framework is unfalsifiable}

A final objection is that any linguistic change after the introduction of generative AI could be attributed to RSN. The framework is designed to prevent this. Its boundary conditions exclude changes without a plausible pathway through machine-generated material (Sect. 3.5); its analytical model defines machine-attributable change relative to a counterfactual baseline (Sect. 3.4); and Section 7.3 states patterns that would restrict or disconfirm it. The framework is therefore selective rather than totalizing. It aims to explain a particular mechanism of semantic reproduction and transformation under computational mediation, not language as such, and it does not replace existing theories of meaning, normativity, social coordination, or semantic change.

\section{Implications and limitations}

\subsection{Implications for AI-mediated societies}

If RSN operates, the diversity of a community's semantic repertoire becomes a collective resource that can be depleted without anyone intending to deplete it, through the aggregation of individually reasonable acts of uptake. We highlight five implications, which are not exhaustive and which we state as proposals rather than conclusions.

First, institutions with high persistence, such as schools and universities, journals, courts, and public administrations, bear particular responsibility, because their uptake is the most consequential (P3). Before machine-generated formulations are codified in rubrics, style guides, templates, or model answers, these institutions can ask where the formulations came from and which alternatives they displace. Given the role \cite{bourdieu1991language} attributes to schooling in legitimating language, educational institutions deserve particular attention. Second, provenance matters for public meaning, not only for authorship: because reproduction leaves no trace in frequency data (P2), knowing what share of circulating material passes through computational channels is a precondition for understanding how a community's semantic environment is maintained, and provenance laundering (Sect. 6.2) means that such knowledge cannot be recovered from documents alone. Third, the correlation of proposals at their source (P4) suggests that diversity in generation, through plural models, diversity-preserving decoding, and alignment procedures sensitive to the range of human perspectives, has value beyond the quality of individual outputs, a point also made by \cite{doi:10.1126/science.adt9819}. Fourth, literacy for AI-mediated communication may need to include a reflexive component: the habit of noticing when a formulation is accepted because it is apt and when it is accepted because it is available. Fifth, because uptake is socially stratified (Sect. 5.3), policies that treat AI-assisted writing only as a matter of individual integrity may overlook that, for some groups, it is a means of access to legitimate language; the question is how to secure that access without making their expression dependent on a narrow standard.

\subsection{Limitations}

The framework has several explicit limits. First, it does not claim that AI independently produces semantic norms, nor that statistical regularity is reducible to normativity. Second, it does not claim that every machine-generated utterance contributes to semantic normalization: outputs may be ignored, rejected, ephemeral, idiosyncratic, or socially inert. Third, the paper is conceptual, and its analytical model is deliberately simplified: it takes the range of alternative formulations as given, groups actors into broad classes, abstracts from network structure, and considers one domain at a time. Fourth, the empirical studies that motivate the framework concern mostly English-language and academic or professional text and reach mostly the lexical level, and the sociological analysis of uptake draws on aggregate adoption data and established theory rather than on direct observation of practice. How uptake unfolds in organizational routines and across social positions requires dedicated study, as does the question of how RSN operates in languages with smaller digital footprints than English, where the relation between ordinary linguistic change and what models propose may differ. Fifth, the framework does not offer a political economy of AI-mediated language or address its legal and regulatory dimensions, although its analysis of uptake power and of the concentration of model development bears on them.

\section{Conclusion}

The question this paper has pursued is not whether language-generating systems understand language, but what happens to the social organization of meaning when their outputs become recurrent inputs into it. We have answered it by naming a mechanism, recursive semantic normalization, a semantic-normative feedback loop through which machine-generated formulations can be incorporated into the practices by which linguistic norms are reproduced and modified; by identifying its unit of analysis, the hybrid semantic system, in which the weak hybridity of systems derived from human language becomes the strong hybridity of their recurrent coupling with human and institutional practice; and by specifying when the mechanism operates, through whom, and with what effects. Three conclusions follow that go beyond the individual arguments.

First, the locus of the philosophical problem shifts. Debates about whether model outputs mean or refer treat the model as the place where semantic questions are settled. If the present analysis is correct, the more consequential semantic facts lie downstream, in the social loop through which formulations are taken up, retained, and made binding. Meaning in AI-mediated societies is therefore less a property that machines have or lack than an outcome that human and institutional practices produce with machine-generated material, and one they could produce differently.

Second, uptake becomes a central object for the social sciences. Whether a machine-generated formulation remains a passing suggestion or becomes a norm depends on routines, organizational imitation, and unequal command of legitimate language. This places generative AI within long-standing sociological concerns with the reproduction of symbolic order and the legitimation of language, especially in institutions such as schools, universities, and public administrations, whose uptake is the most persistent. The same institutions are therefore the places where the course of recursive semantic normalization can most readily be observed and, where necessary, deliberately shaped

Third, the stakes are collective and epistemic rather than only individual. A community can rely justifiably on machine-generated material in every single instance and still lose, through the aggregate of such reliance, part of the diversity of formulations and perspectives from which its accepted views can be criticized, and part of its capacity to know where its own defaults come from. The social epistemology of AI therefore needs, alongside an account of when reliance is justified, an account of how reliance reshapes the epistemic environment in which justification takes place.

These conclusions remain conditional. Most machine-generated language will leave no lasting trace, and existing evidence documents the early stages of the mechanism more firmly than its semantic depth. The value of the framework lies in turning a diffuse concern about AI and language into claims that can be tested: whether machine-attributable change reaches framings and concepts, whether uptake is as stratified as the analysis suggests, and whether machine-derived formulations acquire the marks of norms. If they do, the distinctive problem posed by generative AI will not have been that machines generate language, but that some of the language they generate re-enters, largely unnoticed, the processes through which societies decide what can be said, what counts as saying it well, and what can be known.

\section*{Statements and Declarations}

\paragraph{Funding}
The authors received no specific funding for this work.

\paragraph{Competing interests}
The authors have no relevant financial or non-financial interests to disclose.

\paragraph{Data availability}
Not applicable. No datasets were generated or analysed in this study.

\paragraph{Ethics approval and consent}
Not applicable.

\paragraph{Use of generative AI}
During the preparation of this manuscript, the authors used generative AI to assist with language editing. The authors reviewed, verified, and revised all content, and take full responsibility for the content of the publication.

\bibliographystyle{plainnat}
\bibliography{references}  

\end{document}